\documentclass[graybox]{svmult}

\usepackage{mathptmx}       
\usepackage{helvet}         
\usepackage{courier}        
\usepackage{type1cm}        
\usepackage{makeidx}         
\usepackage{graphicx}        
\usepackage{multicol}        
\usepackage[bottom]{footmisc}

\usepackage{amsmath}

\usepackage{natbib}

\usepackage{verbatim}

\begin{document}

\title*{Strategic behaviour and indicative price diffusion in Paris Stock Exchange auctions}
\author{Damien Challet}
\institute{Damien Challet \at Laboratoire de Math\'ematiques et Informatique pour la Complexit\'e et les Syst\`emes, CentraleSup\'elec, Universit\'e Paris Saclay, France, and Encelade Capital SA, Switzerland\\ \email{damien.challet@centralesupelec.fr}
}
%
%
\maketitle


\abstract{We report statistical regularities of the opening and closing auctions of French equities, focusing on the diffusive properties of the indicative auction price. Two mechanisms are at play as the auction end time nears: the typical price change magnitude decreases, favoring underdiffusion, while the rate of these events increases, potentially leading to overdiffusion. A third mechanism, caused by the strategic behavior of traders, is needed to produce nearly diffusive prices: waiting to submit buy orders until sell orders have decreased the indicative price and vice-versa.}\vspace{2ex}

Research in market micro-structure has focused on the dynamical properties of open markets \citep{ohara,BouchaudFarmerLillo}. However, main stock exchanges have been using auction phases when they open and close for a long time\footnote{London Stock Exchange and XETRA (Germany) recently added a mid-day short auction phase.}. Auctions are known to have many advantages, provided that there are enough participants: for example, auction prices are well-defined, correspond to larger liquidity, and decrease price volatility (and bid-ask spreads) shortly after the opening time and before closing time (see e.g. \cite{pagano2003closing,chelley2008market,pagano2013call}). 

Only a handful of papers are devoted to the dynamics of auction phases, i.e., periods during which market participants may send limit or market orders specifically for the auction.  \cite{boussetta2016role} investigate when fast and slow traders send their orders during the opening auction phase of the Paris Stock Exchange and find markedly different behaviors: the slow brokers are active first, while high-frequency traders are mostly active near the end of auctions. In the same vein, \cite{bellia2016low}
show how and when low-latency traders (identified as high frequency
traders) add or remove liquidity in the pre-opening
auction of the Tokyo Stock Exchange. Accordingly, \cite{yergeau2018machine} finds typical patterns of high-frequency  algorithmic trading in the auctions of XTRA.  \cite{challet2018dynamical} analyze anonymous data from US equities and compute the response functions of the final auction price to the addition or cancellation of auction orders as a function of the time remaining until the auction, which have strikingly  different behaviors in the opening and closing auction phases. Finally, chapter 2 of \cite{lehalle2018market} reports typical daily patterns of matched volume, in a spirit similar to that of the present chapter.

\section{Auctions, data and notations}

The opening auction phase of Paris Stock Exchange starts at 7:15 and ends at 9:00 while the closing auction phase is limited to the period 17:30 to 17:35. The auction price maximises the matched volume. 

From the Thomson Reuters Tick History, we extract  auction phase data for the 2013-04-16 components of the CAC40 index. This database contains all the updates to either the indicative match price or the indicative matched volume in the 2010-08-02 to 2017-04-12 period, which amounts to  8,095,524 data points for the opening auctions and 15,007,048 for the closing auctions. Note that the closing auction phase has about twice as many updates despite being considerably shorter.

For each asset $\alpha$, we denote the indicative price of auction $x\in\{\textrm{open},\textrm{close}\}$ of day $d$ at time $t$ by $\pi_{\alpha,d}^{x}(t)$, the time of auction $x$ by $t^x$ and the auction price by $p_{\alpha,d}^{x}$. Dropping the index $\alpha$ since this paper focuses on a single asset at a time, the $i$-th indicative price change occurs at physical time $t_{i,d}^x$ and its log-return equals $\delta p_{i,d}^x=\log \pi_d^x(t_i)-\log  \pi_d^x(t_{i-1})$. It is useful to work in the time-to-auction (TTA henceforth) time arrow: setting $\tau=t^x-t$,  the log-return between the final auction price and the current indicative is then  $\Delta p_d^x(\tau)=\log p_d^x-\log  \pi_d^x(t)$

Similarly, the indicative matched volume is written as $W_{d}^x(t)$, while the final volume is $V_{d}^x$. Finally, when computing averages over days, since updates occur at random times,  we will use time coarsening by $\delta\tau$ seconds, i.e. compute quantity averages over days within time slices of $\delta \tau $ seconds. 

\begin{figure}
\begin{center}

\includegraphics[width=0.7\textwidth]{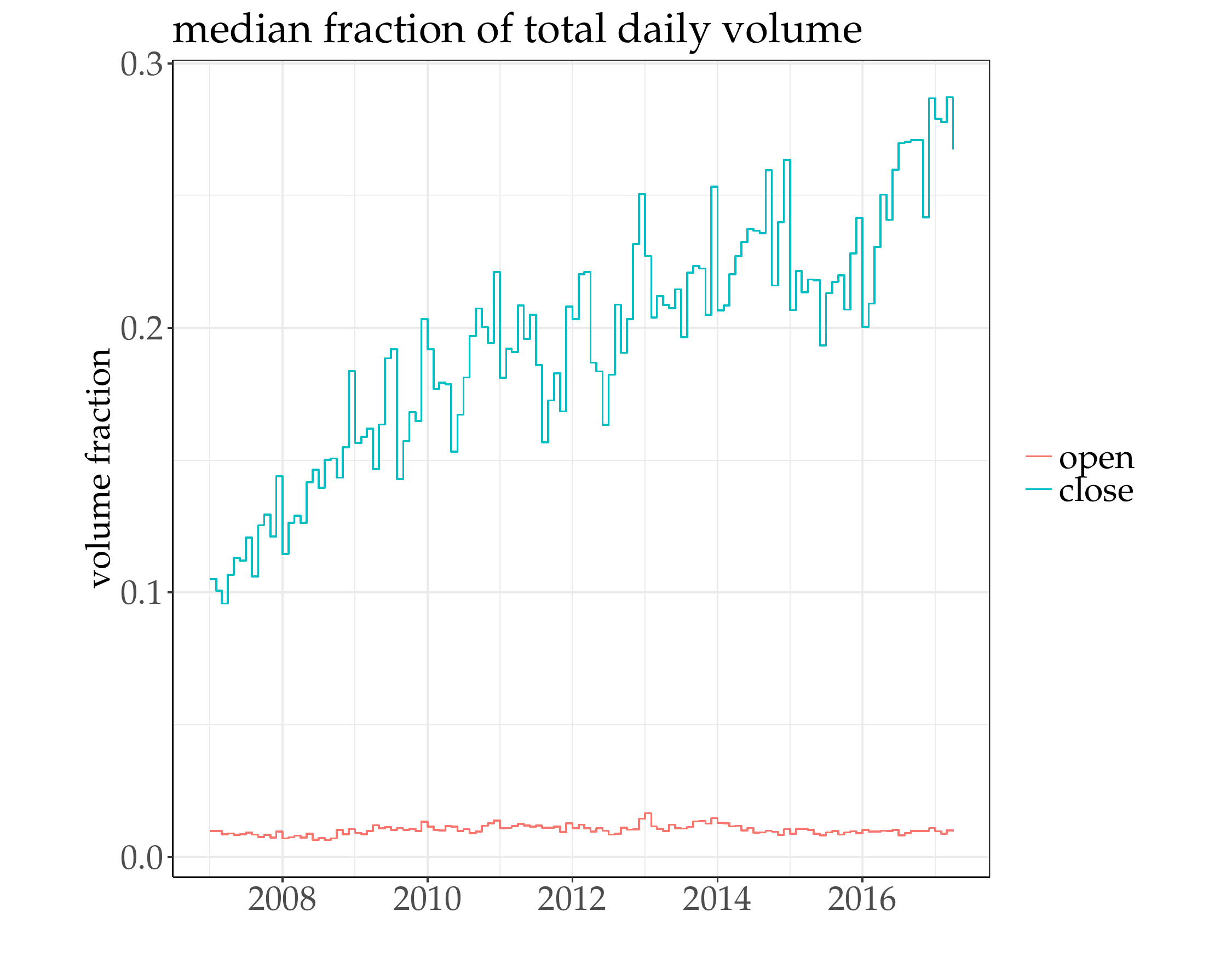}
\caption{Opening and closing fraction of the total daily volume (median computed over all the tickers) since 2007, showing the global increase of the relative importance of the closing auction, but not of the opening auction. Medians over assets of monthly medians for single assets. \label{fig:fracOpenClose}}
\end{center}
\end{figure}

Figure \ref{fig:fracOpenClose} illustrates why auctions deserve  attention: the relative importance of the closing auction volume has more than doubled in the last 10 years. Note that the relative opening auction volume of French equities is quite small (typically around 1\%) and has stayed remarkably constant.

\section{From collisions in event time to diffusion in physical time}

It is useful to consider the price as the position of a uni-dimensional random walker and assume that  each price change is caused by a collision: if collision $i$ shifts the price $p$ by $\delta p_i$, after $N$ collisions the mean square displacement  equals
\begin{equation}
E([\sum_{i=1}^N\delta p_i]^2) \propto N
\label{Edpi2}
\end{equation}

  if the increments $\delta p_i$ are i.i.d, a straightforward consequence of the central limit theorem. This corresponds to standard diffusion. In addition, if the collisions occur at a constant rate $\rho$, then time is homogeneous and $E(t_i)=i\rho$. As we shall see, none of these assumptions is true during auctions, which makes them quite interesting dynamical systems.

\subsection{Event rates}

\begin{figure}
\begin{center}

\includegraphics[width=0.5\textwidth]{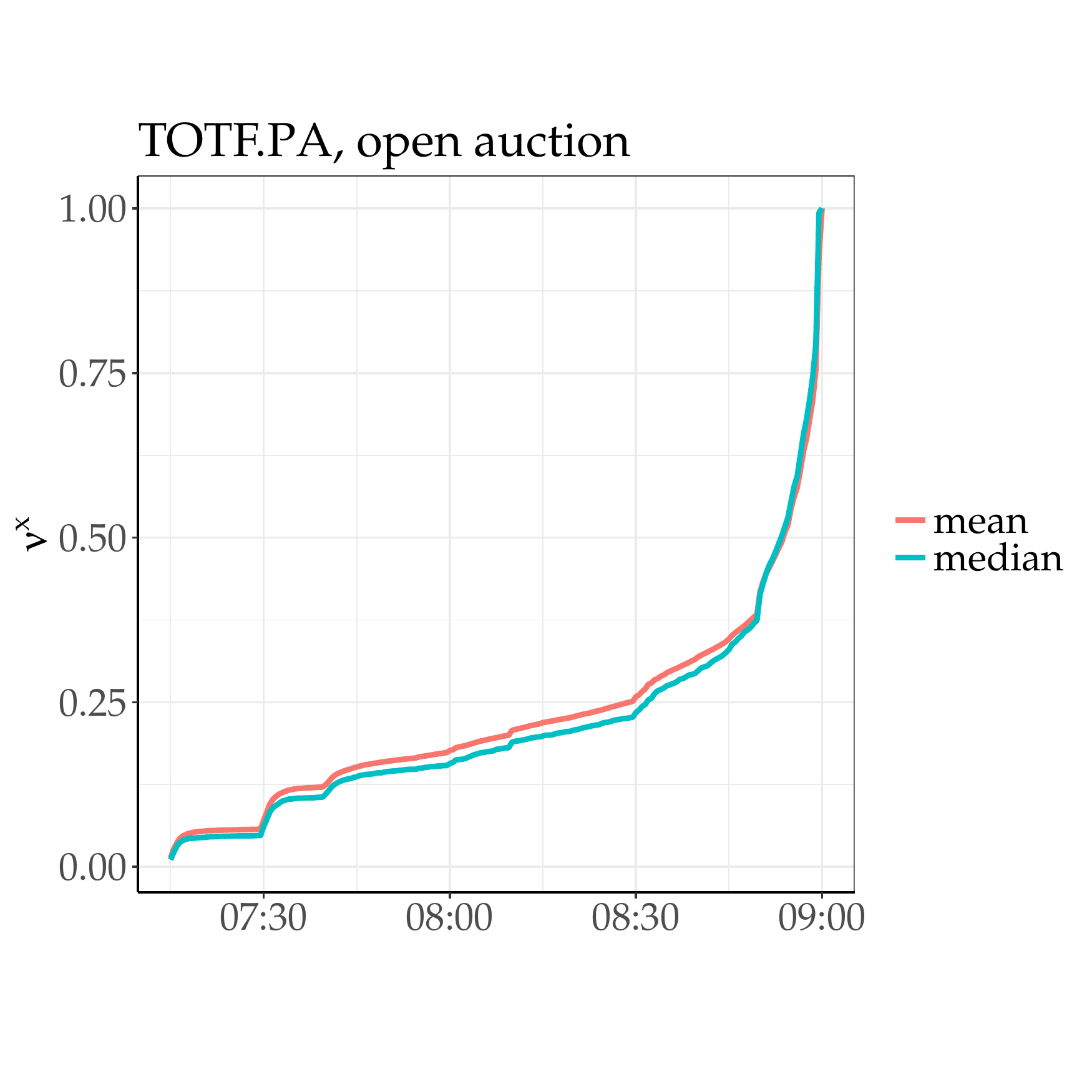}\includegraphics[width=0.5\textwidth]{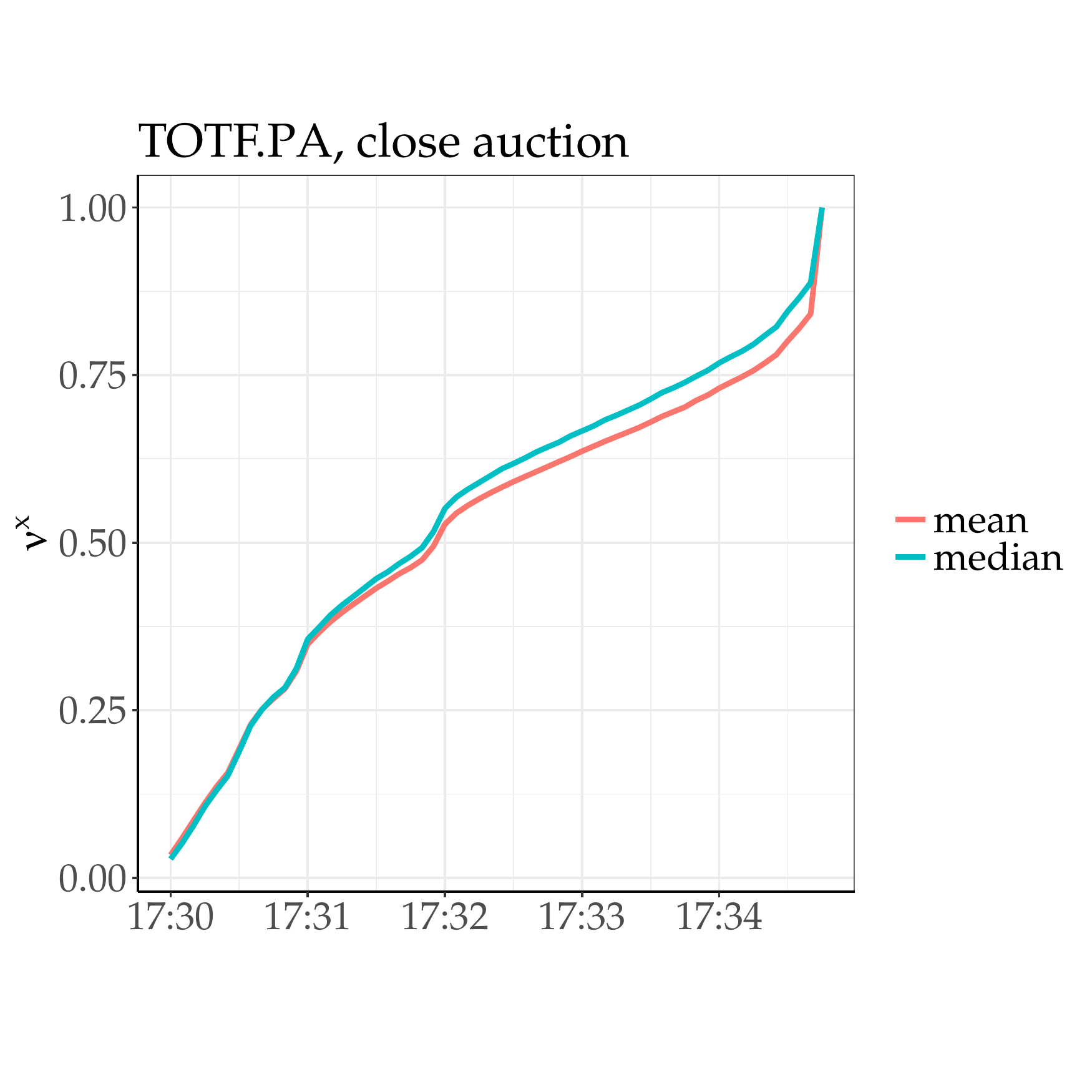}

\includegraphics[width=0.5\textwidth]{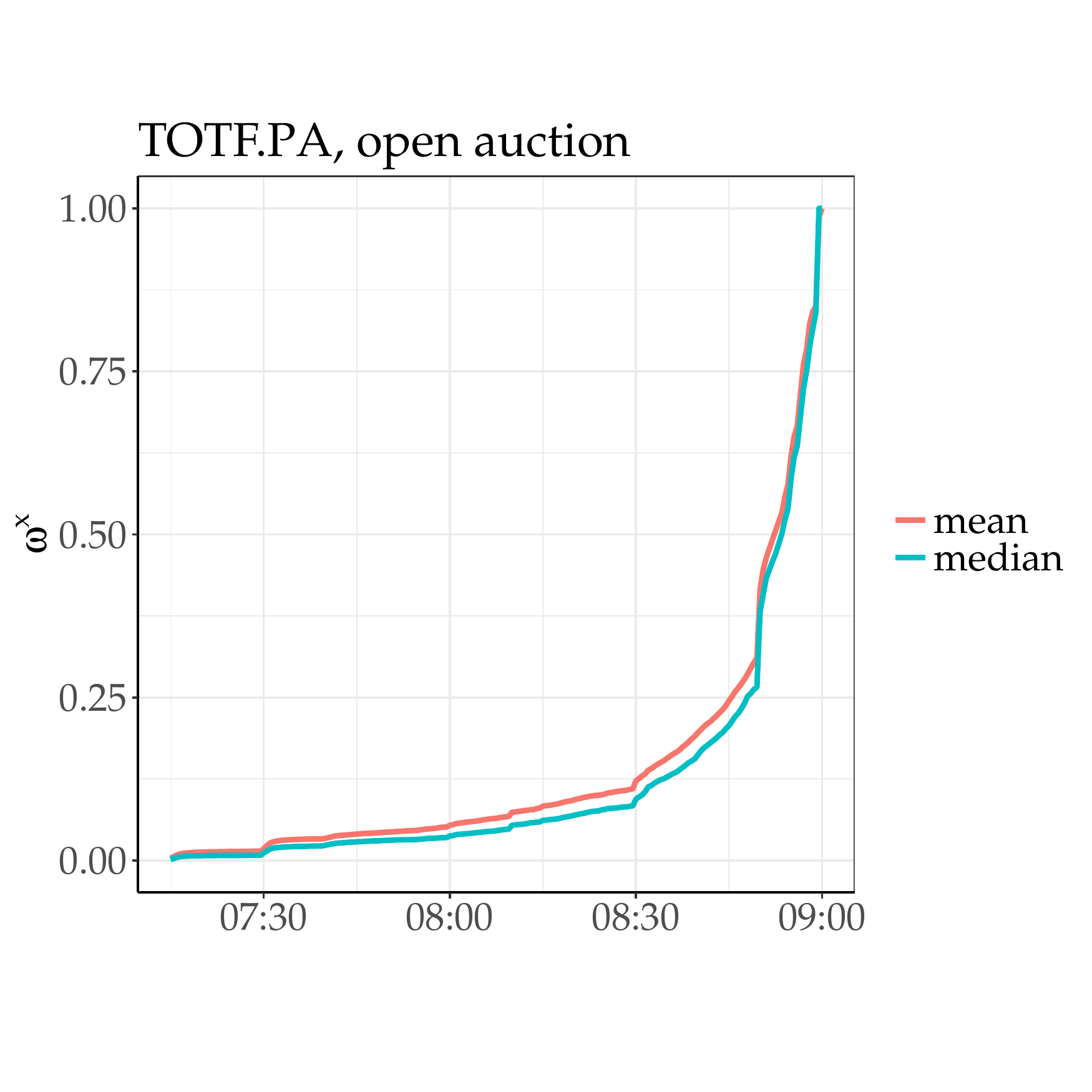}\includegraphics[width=0.5\textwidth]{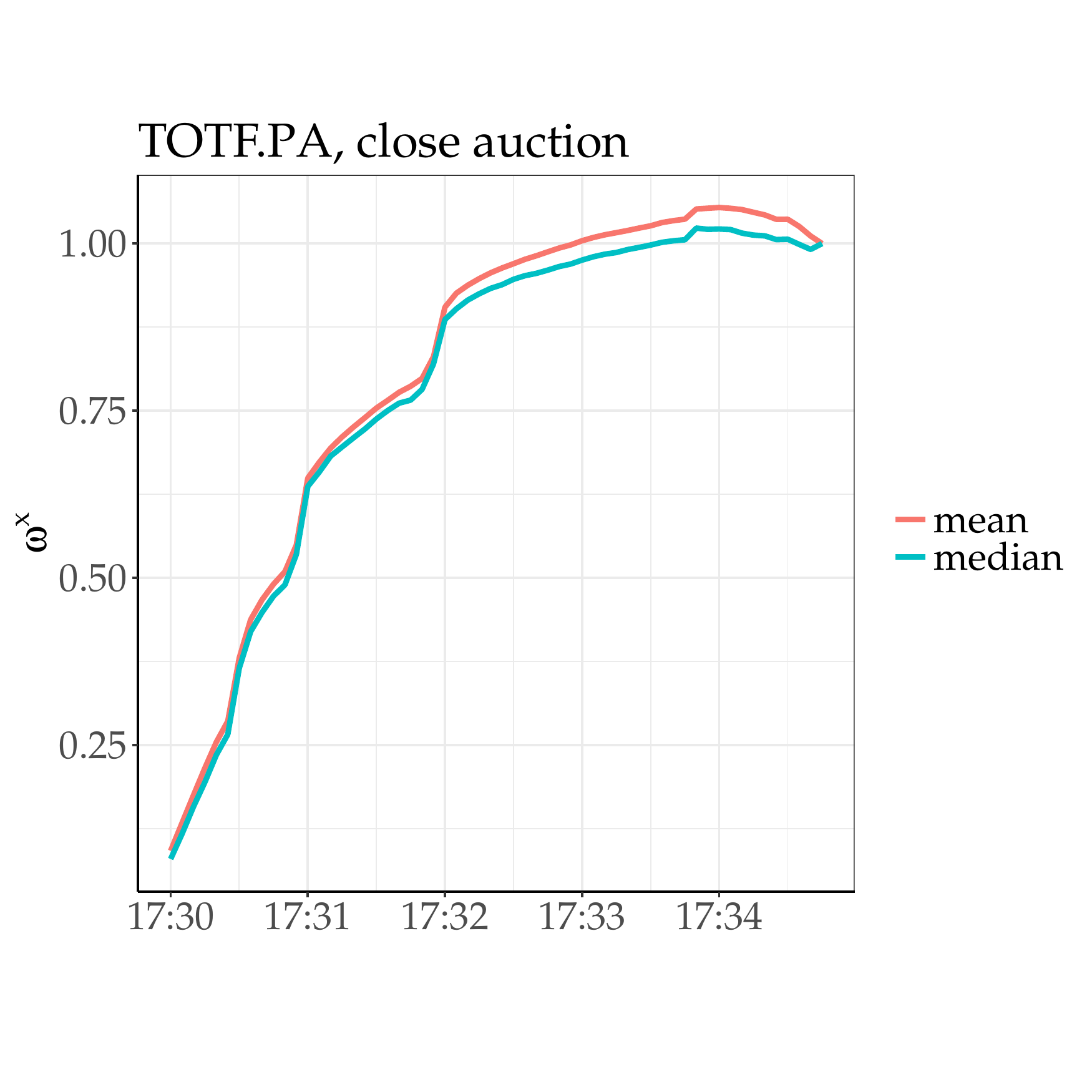}
\caption{Average activity patterns for opening and closing auctions (left and right plots, respectively) for the most active asset (Total). Upper plots: scaled price change events $\nu$ at a function of physical time $t$. Lower plots: scaled indicative volume fraction $\omega$ as a function of $t$. $\delta \tau=30$ seconds for opening auctions and 5 seconds for closing auctions.\label{fig:fractions}}

\end{center}
\end{figure}

In the case of indicative auction prices, the event rate is not constant: the activity  usually increases just before the auction time. This finding is a generic feature of auctions with fixed end time \citep{borle2006timing}, and more generally of human procrastinating nature when faced with a deadline, be it conference registration \citep{alfi2007conference} or paying its fee \citep{alfi2009people}.

 Let us denote by $N_d^x(t)=\sum_{i,~0<t_{i,d}^x\le t} 1$ the number of price events (changes) having occurred up to time $t$ on day $d$ for auction $x$. The activity pattern of day $d$ can be measured by the ratio between the number of events up to time $t$ on day $d$ and the total number of events which happened that day, defined as $\nu_d(t)= N_d^x(t)/N^x_d(t^x)$. The average and median   $\nu(t)=M(\nu_d)(t)$, where $M$ stands for either average or median over days, can be seen in Fig.\ \ref{fig:fractions}. One similarly defines the fraction between the indicative matched volume at time $t$ and the auction volume $\omega(t)=M(W_d^x(t)/V^x_d)$, reported in the same figure. 
 
There are clear peaks of changes for both $\nu$ and $\omega$ at unimaginative physical times such as 7:30, 8:30, etc., and at round minutes and multiples of 30 seconds during the closing auction. This of course denotes a regular behavior of some investors. If each peak is systematically caused by a single trader, there are reasons to think that this regularity does inject information and that it will be exploited by more flexible traders. However, sending orders at the same time as other traders is a rational behavior as it allows one to hide in the crowd, unless one's orders are systematically of the same imbalance sign as the aggregate volume at that time. Thus, from a game theoretical point of view, the emergence of activity peaks is self-organized and stable. Nothing constrains the number of peaks and their locations, which are hence instances of emerging, self-organized conventions. The closing auction being much shorter than the opening one, it is natural that the peaks should appear at round minutes, as this somehow  provides more obvious peak locations than the opening auction. When the closing auction lasts for a much longer time, e.g. for US equities, there  are much fewer price activity peaks \citep{challet2018dynamical}.

The global pattern of price changes and total volume matched clearly differs between both types of auctions. During opening auctions, the price change rate increases much, starting from a low baseline. During closing auctions, the opposite happens: price change activity is first large, slows down during the first 2-3 minutes and then picks up again just before the cut-off time (17:34:45). The average relative matched volume  $\omega(t)$ behaves similarly as $\nu(t)$ during the opening auctions, probably because prices changes are mostly caused by the arrival of new matchable volume, not cancellations. Indeed, half of the open auction events typically happen in the last 10 minutes for most assets, and  half of the volume is matched in the last minutes.  Closing auctions display a different behavior: more than half of the volume is matched during the first minute, and 80\% during the first two minutes. For a few  assets (TOTF.PA, UNBP.PA, for ones), there is a peak of indicative matched volume up to 10\% larger than the auction volume about one minute before the end of the auction; the same behavior is found in US equity markets \citep{challet2018dynamical}.  

\subsection{Activity acceleration}

\begin{figure}
\begin{center}
\includegraphics[width=0.8\textwidth]{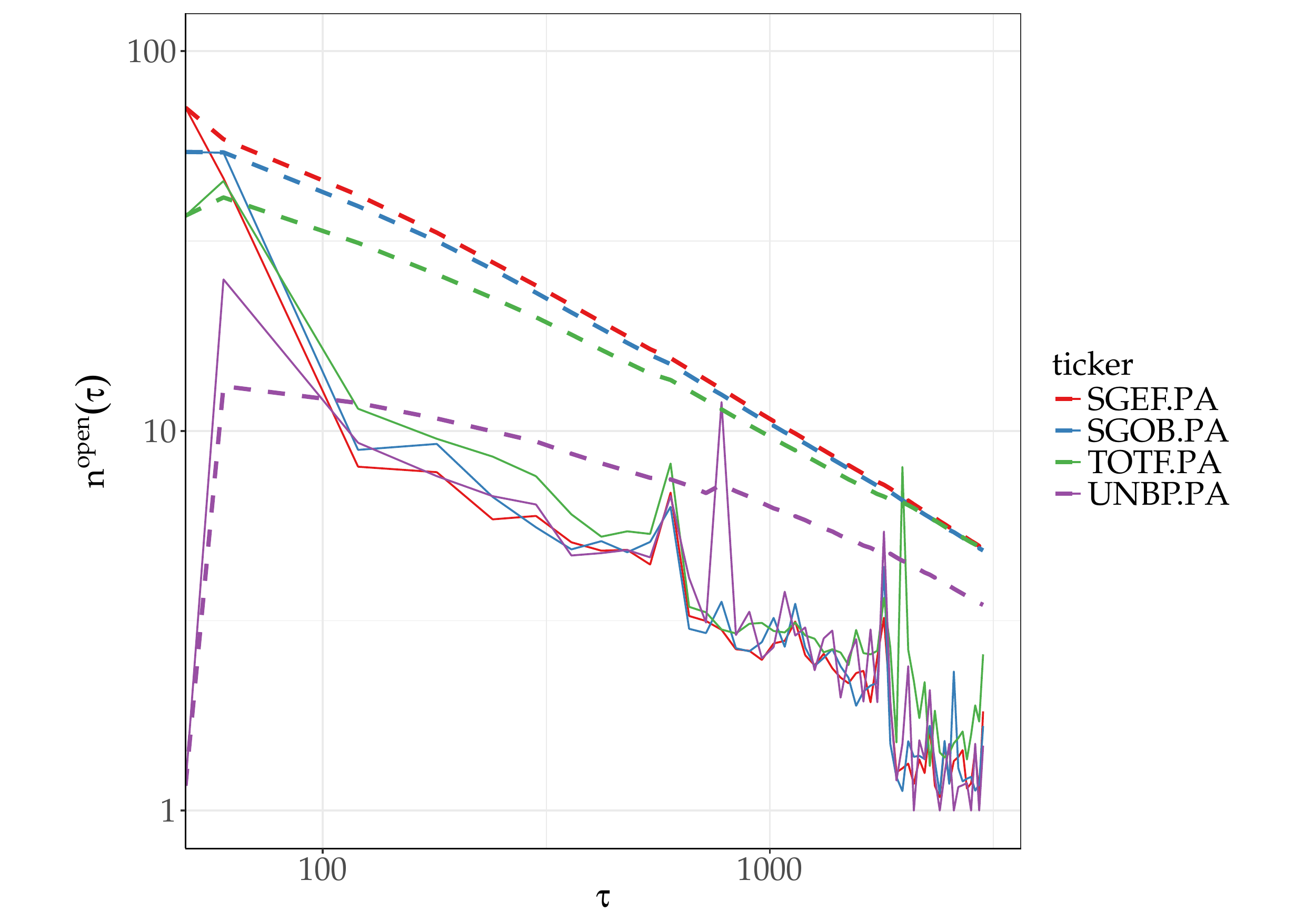}
\end{center}
\caption{Average number of price changes as a function of the time to auction $\tau$, in seconds, for the opening auction. Dashed lines refer to $n_{\small\textrm{smoothed}}$. Time coarsing factor $\delta \tau=60$s. \label{fig:rho}}
\end{figure}

The acceleration pattern of price change rate  follows some regularity. To characterize it in a simpler way, it is useful to work in Time-To-Auction $\tau$ frame. Since the latter reverts the time  arrow, the activity decelerates as a function of $\tau$.  Let us denote the average event rate $\rho^x(\tau)$ so that the expected number of event in the period $\tau$ to $\tau+\delta \tau $ is $n^x(\tau) =E[N_d^x(\tau+ \delta  \tau )-N_d^x(\tau)]=\rho^x(\tau)\delta\tau$.  Figure \ref{fig:rho} shows $n^{\small\textrm{open}}(\tau)$ of several assets, together with the smoothed version of $n^x$, denoted by $n_{\small \textrm{smoothed}}^x=N^x(\tau)/\tau$: if $n^x\propto\tau^{-\beta}$,  so does $n^x_{\small \textrm{smoothed}}$ but with much less noise, which helps assessing the presence of a power-law visually.  We shall drop the $x$ superscript when no confusion is possible.

Assuming that $n(\tau)\propto \tau^{-\beta}$,  we perform a robust linear fit of $\log n(\tau)=cst -\beta \tau$ for $\tau\in[100,300]$ seconds and only keep the fits whose t-statistics associated with $\beta$ is larger than 5. This particular choice of interval for $\tau$ corresponds to a typical period during which the autocorrelation of $\delta p_i$ at one lag is roughly constant (see subsection \ref{ssec:dpi}). In addition, for each asset, we only keep days during which there were at least 50 price changes.

If the typical absolute value of price change $\sigma$ does not depend on $\tau$ and is still i.i.d., Eq.\ (\ref{Edpi2}) becomes 
\begin{align}
E([\sum_{i=1}^N\delta p_i]^2) &=\sum_{i:\ {t_i}<=\tau}E(\delta p_i)^2\propto\sigma\tau^{1-\beta}
\label{eq:Edpi2_ntau}
\end{align}

hence the Hurst exponent in $\tau$ time, denoted by $h$, equals $(1-\beta)/2$: the price change rate influences the diffusive pattern in a simple way, given the above approximations. It is worth noting at this juncture that in the normal time frame the price is overdiffusive if $\sigma$ does not depend on $\tau$ and if $\beta>0$, i.e., if the rate of price changes increases near the auction end time and the Hurst exponent in the normal time arrow is $H=(1+\beta)/2$.

\subsection{Typical price change}

\begin{figure}
\begin{center}
\includegraphics[width=0.8\textwidth]{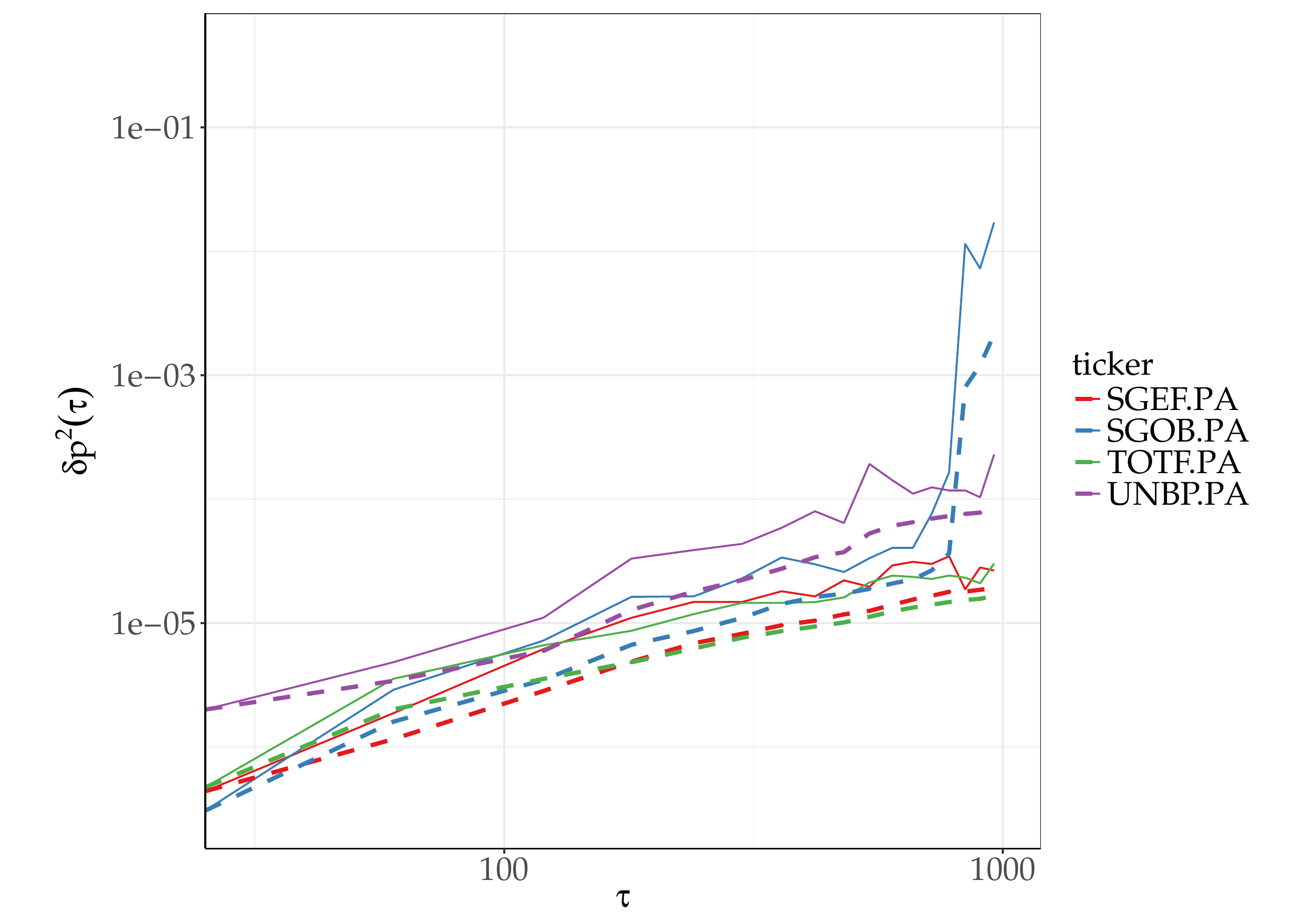}
\end{center}
\caption{Average scale of the log price increment as a function of the time to auction $\tau$, in seconds, for the opening auction. Dashed lines refer to the smoothed quantity. \label{fig:dp2_vs_tau}}
\end{figure}

When the indicative price changes, it jumps to the next non-empty tick of the auction order book. Thus, the typical indicative price change reflects the density of the latter, which increases as the auction time nears. As a consequence, the typical price change magnitude $\sigma$ is not constant but decreases near the auction end time, or equivalently increases as a function of $\tau$. Once again, for opening auctions, we find an approximate power-law relationship $\sigma(\tau)\propto \tau^\alpha$ (see Fig.\ \ref{fig:dp2_vs_tau}). We apply the same method as for $n(\tau)$ to estimate $\alpha$:  we only keep days during which there were at least 50 price changes for a given asset;  robust fits  of $\log \delta p (\tau)=cst +\alpha \tau$ for $\tau\in[100,300]$ are carried out. Only fits whose t-statistics associated with $\alpha$ are larger than 5 are kept. 

\subsection{Diffusive properties of  indicative prices}
\label{ssec:dpi}

It is easy to see why the increase of activity and decrease of the typical magnitude of price changes have antagonistic and purely mechanistic effects on the diffusive properties of the indicative auction price in the simplest case: neglecting the autocorrelations and cross-correlations of both $n(\tau)$ and $\delta p_i$, Eq.\ (\ref{eq:Edpi2_ntau}) becomes indeed
\begin{align}
E\left(\Delta p^2\right)(\tau)&\simeq  \sum_{\tau'<=\tau}E(n(\tau') \delta p^2(\tau'))\propto \tau ^{h_0} \label{eq:Edpi2_n_sigma_tau_1}
\\
& \simeq\sum_{\tau'<=\tau}E(n(\tau'))E( \delta p^2(\tau'))\propto \tau^{1+\alpha-\beta}=\tau^{h_0^{(\alpha\beta)}},
\label{eq:Edpi2_n_sigma_tau_2}
\end{align}

\begin{figure}
\begin{center}
\includegraphics[width=0.5\textwidth]{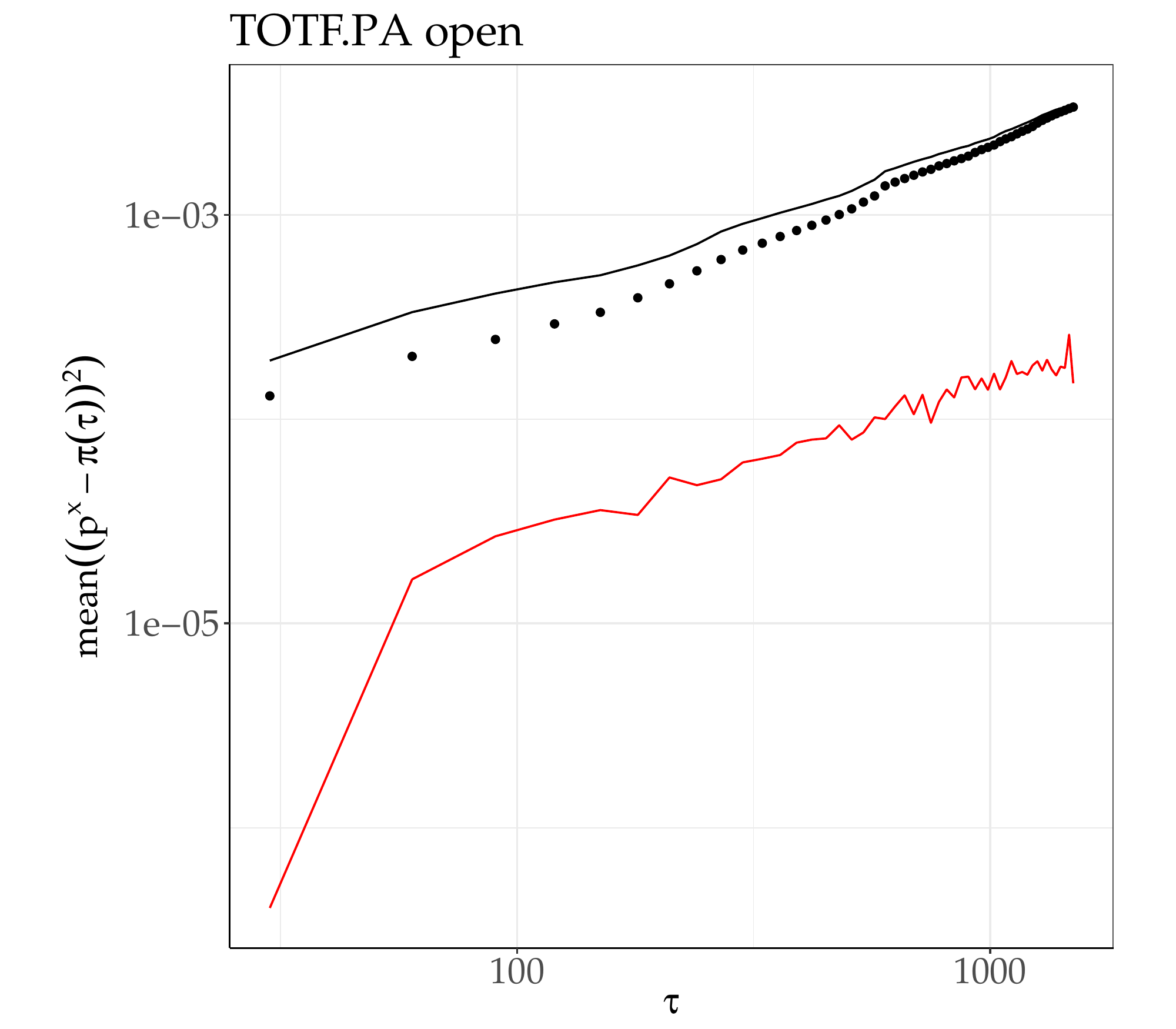}\includegraphics[width=0.5\textwidth]{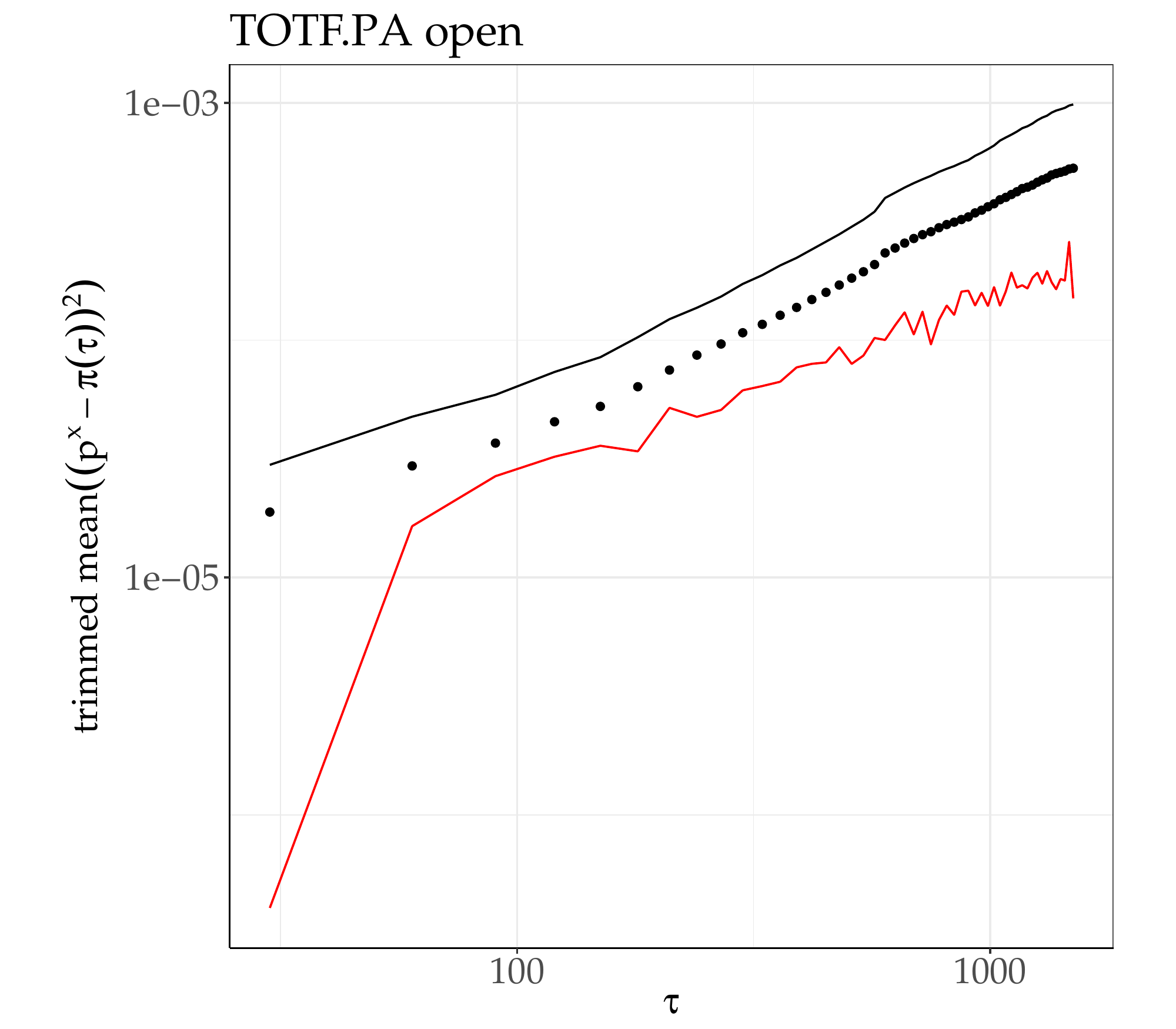}
\end{center}
\caption{Average square difference between the auction price and the indicative price $\tau$ seconds before the opening auction. Continuous red lines (bottom of the figure) refer to $E(\Delta p^2)(\tau)$, The upper black continuous line is $ \sum_{\tau'<=\tau}[n(\tau') \delta p^2(\tau')]$, and the black dots are $ \sum_{\tau'<=\tau}E[n(\tau')]E[ \delta p^2(\tau')]$. Left plot: plain averages over all values of  $\delta p_i$; right plot: trimmed means where the 20\% largest (in absolute value) $\delta p_i$ for each day and each time slice of $\delta  \tau=30$ seconds have been removed in the computation of the averages of quantities based on $\delta p_i$.\label{fig:Deltap2_vs_tau}}
\end{figure}

The first approximation assumes that all $\delta p_i$ within a time slice are i.i.d, while the second one assumes no correlation between $n$ and $\delta p^2$. The relative merits of both approximations can be assessed in Fig.\ \ref{fig:Deltap2_vs_tau}. The first approximation corresponds to the continuous black line and the second one to the black dots. Both curves are close together; however neglecting the dependence between $n$ and $|\delta p|$ underestimates the typical magnitude of $\Delta p$. The same figure makes it clear that something is wrong even in the first approximation, as $ \sum_{\tau'<=\tau}E(n(\tau') \delta p^2(\tau'))$ is about 10 times larger than $E(\Delta p^2)$.  This discrepancy is mainly due to the bouncing behavior of $\pi$ for large $\tau$: a large $\delta p_i$ is typically followed by large $\delta p_{i+1}$ of opposite sign, which  inflates $E(\delta p^2)$ and does not correspond to significant price change as the latter reverts immediately to a value close to that before event $i$. This is why trimmed means, which removes a given fraction of the largest $\delta p_i$ for each time slice and each day, decrease much this discrepancy. The latter is also due in part to a simple strategic behavior: during the auction phase, negative indicative price change triggers the sending of buy orders and vice-versa, causing an intrinsically smaller than expected $\Delta p (\tau)^2$ (see below for a more detailed discussion).

\begin{figure}
\begin{center}

\includegraphics[width=0.5\textwidth]{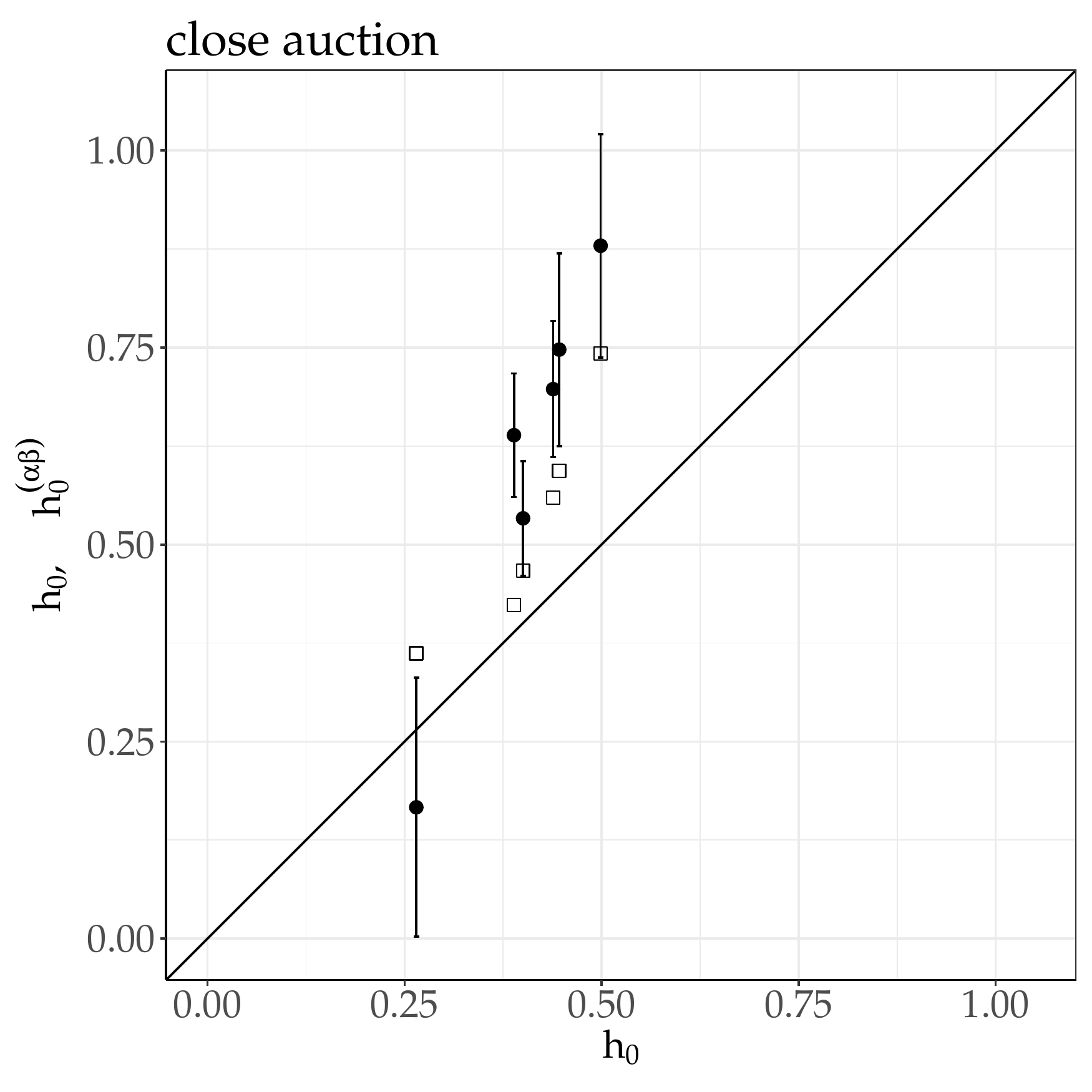}
\caption{Hurst exponents $h_0$ and $h_0^{(\alpha\beta)}$ versus the actual Hurst exponent for the 6 assets of the CAC40 that yield good power-law fits of both $\sigma^2\propto \tau^\alpha$  and $n\propto\tau^{-\beta}$; open auction, $\delta \tau=60[s]$; Time-To-Auction arrow. Error bars correspond to one standard deviation. When no error bar is visible, the error is at most as large as the symbol. \label{fig:Hsynth_vs_H}}
\end{center}
\end{figure}

Let us now compare the TTA Hurst exponents of the above quantities, plotted in Fig.\ \ref{fig:Hsynth_vs_H} for the 6 stocks whose fits of both $\alpha$ and $\beta$ are deemed significant. Two features stand out. First, $h_0$ overestimates $h$, even when accounting for the fairly large error bars. This implies that the dynamics caused by the interplay between typical price change shrinking and the acceleration of the activity is more subtle than the simple approximation above. In fact, interestingly, $h_0$ also overestimates the Hurst $h_0$ exponent: this emphasizes the fact that the  $\delta p_i$ are not i.i.d. 


\begin{figure}
\begin{center}
\includegraphics[width=0.5\textwidth]{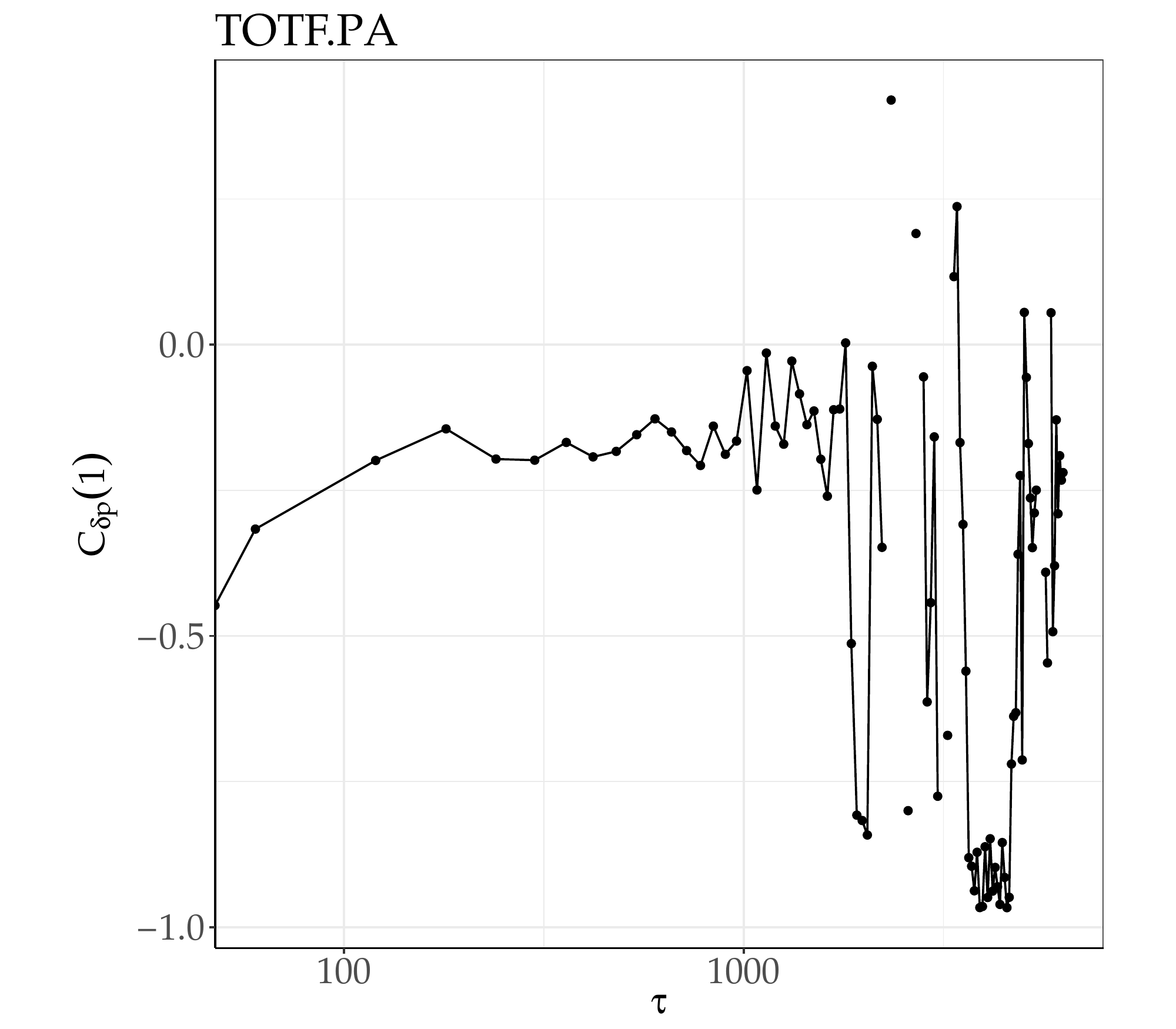}
\caption{Autocorrelation between two consecutive price changes within time slices of $\delta \tau=60$ seconds,   averaged over all days for TOTF.PA. \label{fig:dp_acf}}
\end{center}
\end{figure}

Indeed, in practice,  even linear autocorrelation of both $\rho(\tau)$ and $\delta p_i$ and the cross-correlation between them are  not negligible. Let us focus on the autocorrelation of $\delta p_i$, denoted by $C_{\delta p}(\delta i)$. For each time slice $[\tau,\tau+\delta \tau[$, we average $C_{\delta p,d}(1)$ over all the days for a given asset. Figure \ref{fig:dp_acf} plots this quantity versus $\tau$ for TOTF.PA, the most active asset in our dataset. Generally, $C_{\delta p}(1)<0$; even more, it becomes more and more negative near the auction time, i.e., for small $\tau$. Since the price changes become relatively smaller in that limit, this reflects a purposeful bounce of the indicative auction price between two close price ticks; the large negative autocorrelation points to strategic behavior, by which traders try to decrease the immediate impact of their auction orders by submitting their orders after other orders of the opposite sign (hence to hide their actions); in fact, the autocorrelation of the sign of $\delta p$, $C_{\rm{sign}\,\delta p}(1)$ is even smaller than  $C_{\delta p}(1)$ for small $\tau$. For large $\tau$, this auto-correlation also tends to have very small values, which is reinforced by the fact that an outstandingly large $\delta p_i $ is often followed by a similarly large $\delta p_{i+1}$ of opposite sign. Thus strategic behavior is more common for small $\tau$.

When $C_{\delta p}(1)$ does not depend on $\tau$, it only modifies the prefactor of $\tau$ in Eq.\ (\ref{eq:Edpi2_n_sigma_tau_1}) by a factor of the order $\frac{1+C_{\delta p}(1)}{1-C_{\delta p}(1)}$, not the Hurst exponent, and thus explains in part the discrepancy between $E(\Delta p^2)(\tau)$ and $\sum_{\tau'=1}^N E[n(\tau')\delta p(\tau')^2]$. The dependence of  $C_{\delta p}(1)<0$ on $\tau$ modifies the apparent Hurst exponent in a nontrivial way. This is why we measured $h$ for $\tau\in[100,300]$, i.e., in a region where $C_{\delta p}(1)<0$ is the most constant. 

\section{Discussion}

Indicative auction prices display non-trivial properties due in part to the antagonistic effects of both the acceleration of activity and the reduction of the typical price change magnitude.  However, the indicative price is much less over-diffusive than what these two effects alone imply. In other words, the deviation from purely mechanistic effects points to a more subtle dynamics. This makes sense, as the traders have a clear incentive to minimize their easily detectable impact. Their strategic behavior results in often  alternatively positive and negative indicative price changes, i.e., in a clearly anti-correlated price changes. Quite tellingly, this negative auto-correlation is more and more pronounced as the auction end nears.

So far, we have used a basic data type, which nevertheless has a rich behavior. More detailed data, such as data from the auction book, will allow us to characterize order strategic placement, the evolution of the average auction book density and the price impact of new orders and order cancellations much before the auction time, in the spirit of the response function of \cite{challet2018dynamical}, but accounting for both the volume of new auction orders and their immediate impact on the auction order book.

\bibliographystyle{plainnat}
\bibliography{biblio}

\end{document}